\documentclass[prl,twocolumn,superscriptaddress]{revtex4-1}

\usepackage{microtype}
\usepackage[normalem]{ulem}
\usepackage{color}
\usepackage{graphicx}
\usepackage{amssymb,amsmath,braket,cancel}
\usepackage{siunitx}
\usepackage{commath} 
\usepackage[unicode=true]{hyperref}
\usepackage[doipre={doi:}]{uri}

\begin{document}

\title{Exotic Looped Trajectories of Photons in Three-Slit Interference}
\author{Omar S. Maga\~{n}a-Loaiza}
\email{omar.maganaloaiza@rochester.edu}
\affiliation{The Institute of Optics, University of Rochester, Rochester, New York 14627, USA}
\author{Israel De Leon}
\email{ideleon@itesm.mx}
\thanks{OSML and IDL contributed equally to this work.}
\affiliation{School of Engineering and Sciences, Tecnol\'{o}gico de Monterrey, Monterrey, Nuevo Leon 64849, Mexico}
\affiliation{Department of Physics and Max Planck Centre for Extreme and Quantum Photonics, University of Ottawa, Ottawa, Ontario, K1N 6N5, Canada.}
\author{Mohammad Mirhosseini}
\affiliation{The Institute of Optics, University of Rochester, Rochester, New York 14627, USA}
\author{Robert Fickler}
\affiliation{Department of Physics and Max Planck Centre for Extreme and Quantum Photonics, University of Ottawa, Ottawa, Ontario, K1N 6N5, Canada.}
\author{Akbar Safari}
\affiliation{Department of Physics and Max Planck Centre for Extreme and Quantum Photonics, University of Ottawa, Ottawa, Ontario, K1N 6N5, Canada.}
\author{Uwe Mick}
\affiliation{Max Planck Institute for the Science of Light, Guenther-Scharowsky-Str. 1, Erlangen, Germany}
\author{Brian McIntyre}
\affiliation{The Institute of Optics, University of Rochester, Rochester, New York 14627, USA}
\author{Peter Banzer}
\affiliation{Department of Physics and Max Planck Centre for Extreme and Quantum Photonics, University of Ottawa, Ottawa, Ontario, K1N 6N5, Canada.}
\affiliation{Max Planck Institute for the Science of Light, Guenther-Scharowsky-Str. 1, Erlangen, Germany}
\author{Brandon Rodenburg}
\affiliation{Quantum Information Science Group, MITRE, 200 Forrestal Rd. Princeton, NJ 08540, USA}
\author{Gerd Leuchs}
\affiliation{Department of Physics and Max Planck Centre for Extreme and Quantum Photonics, University of Ottawa, Ottawa, Ontario, K1N 6N5, Canada.}
\affiliation{Max Planck Institute for the Science of Light, Guenther-Scharowsky-Str. 1, Erlangen, Germany}
\author{Robert~W.~Boyd}
\affiliation{The Institute of Optics, University of Rochester, Rochester, New York 14627, USA}
\affiliation{Department of Physics and Max Planck Centre for Extreme and Quantum Photonics, University of Ottawa, Ottawa, Ontario, K1N 6N5, Canada.}

\begin{abstract}
{The validity of the superposition principle and of Born's rule are well-accepted tenants of quantum mechanics. Surprisingly, it has recently been predicted that the intensity pattern formed in a three-slit experiment is seemingly in contradiction with the predictions of the most conventional form of the superposition principle when exotic looped trajectories are taken into account. However, the probability of observing such paths is typically very small and thus rendering them extremely difficult to measure. In this work, we confirm the validity of Born's rule and present the first experimental observation of these exotic trajectories as additional paths for the light
by directly measuring their contribution to the formation of optical interference fringes. We accomplish this by enhancing the electromagnetic near-fields in the vicinity of the slits through the excitation of surface plasmons. This process effectively increases the probability of occurrence of these exotic trajectories, demonstrating that they are related to the near-field component of the photon's wavefunction.}
\end{abstract}

\maketitle


The phenomenon of interference has been recognized as ``the only mystery'' of quantum mechanics~\cite{Mandel:1999kba}. The enormous interest and history of this fundamental effect can be traced back to the two-slit experiment devised by Thomas Young in the early 19th century. Young's experiment is conceptually the simplest method for demonstrating the superposition principle, as the appearance of interference fringes in the far-field is unexplainable unless it is understood that the particle seemingly travels through both slits simultaneously. Such an experiment, originally performed with light, has since been conducted on particles ranging from individual photons, neutrons, and atoms, to large molecules consisting of dozens of atoms~\cite{Greenberger:1993kba}. As the superposition principle lies at the core of quantum physics, many of its counterintuitive features such as entanglement, non-locality, wave-particle duality, and delayed-choice concepts can be demonstrated or tested using a two-slit system~\cite{Shadbolt:2014ip, Scully:1991if, Kim:2000un, Schouten:2005bl, Kocsis:2011jg, Menzel:2012km, Bolduc:1664716, Mahler:2016kn}.

The standard interpretation of the two-slit experiment is given by solving the wave equation for an initially prepared complex wavefunction, $\psi$. For example, if $\psi_A$ represents the wavefunction at the detector for a photon emerging from slit $A$, and $\psi_B$ is the wavefunction for a photon emerging from slit $B$, then the implementation of the superposition principle is to assume that the wavefunction is a superposition of the different paths given by ${\psi_{AB} = \psi_A + \psi_B}$. The probability of detection is given by Born's rule as
\begin{equation}
P_{\rm AB} \equiv \abs{\psi_{\rm AB}}^2 
    = P_{\rm A} + P_{\rm B} + \left(\psi^*_{\rm A}\psi_{\rm B} + \psi_{\rm A}\psi^*_{\rm B}\right),
	\label{eq:born}
\end{equation}
where $P_{\rm A} = \abs{\psi_{\rm A}}^2$ and $P_{\rm B} = \abs{\psi_{\rm B}}^2$. From this equation it is clear that the outcome of the two-slit experiment is given by the sum of outcomes from each slit alone, plus an additional interference term.

Due to the inherent structure of any wave theory, Born's rule always bounds the complexity of any effect involving superpositions of an arbitrary number of wavefunctions to a sum of terms denoting the interference between pairs of wavefunctions~\cite{Sorkin1994}. For instance, in accordance with Born's rule, the interference pattern obtained in a three-slit experiment can be described by the following probabilities 
\begin{equation}
    P_{\rm ABC} = P_{\rm AB} + P_{\rm BC} + P_{\rm AC} - P_{\rm A} - P_{\rm B} - P_{\rm C}.
\end{equation}
Note that this expression does not include a probability term that involves three slits, but is entirely described by probabilities involving only one and two slits. Any possible contribution from higher-order interference terms (i.e., a path involving the three slits) has been quantified by the so-called Sorkin parameter~\cite{Sorkin1994,Sinha:2010kn}
\begin{equation}
    \epsilon = P_{\rm ABC} - P_{\rm AB} - P_{\rm BC} - P_{\rm AC} + P_{\rm A} + P_{\rm B} + P_{\rm C},
    \label{eqn:epsilon}
\end{equation}
which should be identically zero if only the direct paths through the three individual slits are considered. Sinha~{\it et al.}~\cite{Sinha:2010kn} showed that $\epsilon$ can be evaluated experimentally by making a set of measurements for each term in Eq.~\eqref{eqn:epsilon}.

Although it might appear that the measurement of a non-zero $\epsilon$ implies a clear violation of quantum mechanics~\cite{Sinha:2010kn},  De Raedt~{\it et al.} demonstrated by numerically solving Maxwell's equations that a non-zero value of $\epsilon$ can exist without such violation~\cite{Raedt:2012un}. Later it was found that this result is a consequence of the presence of exotic looped trajectories of light (e.g. red curve in Fig.~\ref{fig:fig1}a) that arise in the Feynman path integral formulation with extremely low probability of occurrence \cite{Sawant:2014eg}.  This interpretation was subsequently shown to agree with the exact numerical solution of the wave equation~\cite{Sinha:2015dk}.

In this work we demonstrate that looped trajectories of photons are physically due to the near-field component of the wavefunction, which leads to an interaction among the three slits. As such, it is possible to increase the probability of occurrence of these trajectories by controlling the strength and spatial distribution of the electromagnetic near-fields around the slits. By a proper control of the conditions in a three-slit experiment, we successfully demonstrate a dramatic increase of the probability of photons to follow looped trajectories, and present the first successful measurement of a non-zero value of $\epsilon$.

\vspace{-0.3 cm}
\subsection*{Origin of the looped trajectories of photons}
\vspace{-0.3 cm}
Under the scalar wave approximation, the propagation of light is described by the Helmholtz equation
\begin{equation}
	\left(\nabla^2 + k^2\right)\psi(\mathbf r) = 0,
    \label{eqn:Helmholtz}
\end{equation}
subject to the boundary conditions specifying the physical setup. This equation can be solved by computing the propagation from any point
$\mathbf{r_{\rm 1}}$ to any other point $\mathbf{r_{\rm 2}}$ via the Green's function kernel, which according to Rayleigh-Sommerfeld theory is given by
\begin{equation}
    K(\mathbf{r_{\rm 1}, r_{\rm 2}}) = \frac{k}{2\pi i}~\frac{e^{ik~\abs{\mathbf{r_{\rm 1}-r_{\rm 2}}}}}{\abs{\mathbf{r_{\rm 1}-r_{\rm 2}}}}\chi,
\end{equation}
where $\chi$ is an obliquity factor \cite{Goodman:Book:1968}. This equation satisfies Eq.~\eqref{eqn:Helmholtz} and the Fresnel-Huygens principle in the form of the following propagator relation
\begin{equation}
    K(\mathbf{r_{\rm 1},r_{\rm 3}}) = \int \dif\mathbf{r_{\rm 2}} K(\mathbf{r_{\rm 1},r_{\rm 2}}) K(\mathbf{r_{\rm 2},r_{\rm 3}}).
    \label{eqn:Fresnel-Huygens}
\end{equation}
If one repeatedly applies Eq.~\eqref{eqn:Fresnel-Huygens}, the path-integral formulation of the propagation kernel is obtained in the form~\cite{Feynman:book:1965}
\begin{equation}
    K(\mathbf{r_{\rm 1},r_{\rm 2}}) = \int \mathcal D[x(\mathbf s)]\exp\left(ik\int\dif\mathbf s\right),
    \label{eqn:PathIntegral}
\end{equation}
where $\int \mathcal D[x(\mathbf s)]$ is the functional integration over paths $x(\mathbf s)$. The boundary conditions can be included by restricting the possible paths $x(\mathbf s)$. If one is concerned only with diffraction from slits in a single plane, then Eq.~\eqref{eqn:PathIntegral} can be perturbatively expanded as~\cite{Sawant:2014eg}
\begin{equation}
    K = K_1 + K_2 + K_3 + \cdots,
    \label{eqn:Ksum}
\end{equation}
where $K_n$ represents the $n$th application of Eq.~\eqref{eqn:Fresnel-Huygens} and each integration is carried over the plane containing the slits~\cite{SuppMat}.

Solving the wave equation taking $K = K_1$ is equivalent to considering only direct paths, such as the paths in Fig.~\ref{fig:fig1}b. These paths propagate from the source and through one of the slits to the detector. We call these wavefunctions $\psi_{\rm A}$, $\psi_{\rm B}$ and $\psi_{\rm C}$. The higher-order terms in Eq.~\eqref{eqn:Ksum} are responsible for the looped trajectories of photons that propagate from the source to a slit, and {\it to at least one other slit} before propagating to the detector (see Fig.~\ref{fig:fig1}c). It follows that the wavefunction of a photon passing through the three slits is given by
\begin{equation}
    \psi_{\rm ABC} = \psi_{\rm A} + \psi_{\rm B} + \psi_{\rm C} + \psi_{\rm L},
	\label{eq:super}
\end{equation}
where $\psi_{\rm L}$ represents the contribution of the looped trajectories to the wavefunction $\psi_{\rm ABC}$. Note that in general $\epsilon$, as defined by Eq.~\eqref{eqn:epsilon}, is not zero because of the existence of these looped trajectories. Thus, the presence of looped paths leads to an apparent deviation of the superposition principle~\cite{Sawant:2014eg,SuppMat}. 

\subsection*{Increasing the probability of occurrence of the looped trajectories of photons}
The conclusion that $\psi_{\rm ABC}$ is not simply the superposition of the wavefunctions $\psi_{\rm A}$, $\psi_{\rm B}$, and $\psi_{\rm C}$ is a consequence of the actual boundary conditions in a three-slit structure. Changing the boundary conditions affects the near-field components around the slits, but it typically does not affect the far-field distribution because of the short range extension of the near fields~\cite{Kowarz1995}. As shown below, the looped trajectories of photons are physically due to the near-field components of the wavefunction. Therefore, by controlling the strengths and the spatial distributions of the near-fields around the slits, it is possible to drastically increase the probability of photons to undergo looped trajectories, thereby allowing a straightforward visualization of their effect in the far-field interference pattern. To demonstrate this phenomenon, a three-slit structure was designed such that it supports surface plasmons, which are strongly confined electromagnetic fields that can exist at the surface of metals~\cite{Raether:book:1988,Barnes:nature:2003}. The existence of these surface waves results in near fields that extend over the entire region covering the three slits~\cite{Gay2006,Schouten2005}, thereby increasing the probability of looped trajectories.

\begin{figure*}[t]
\centering
\includegraphics[width=1\textwidth]{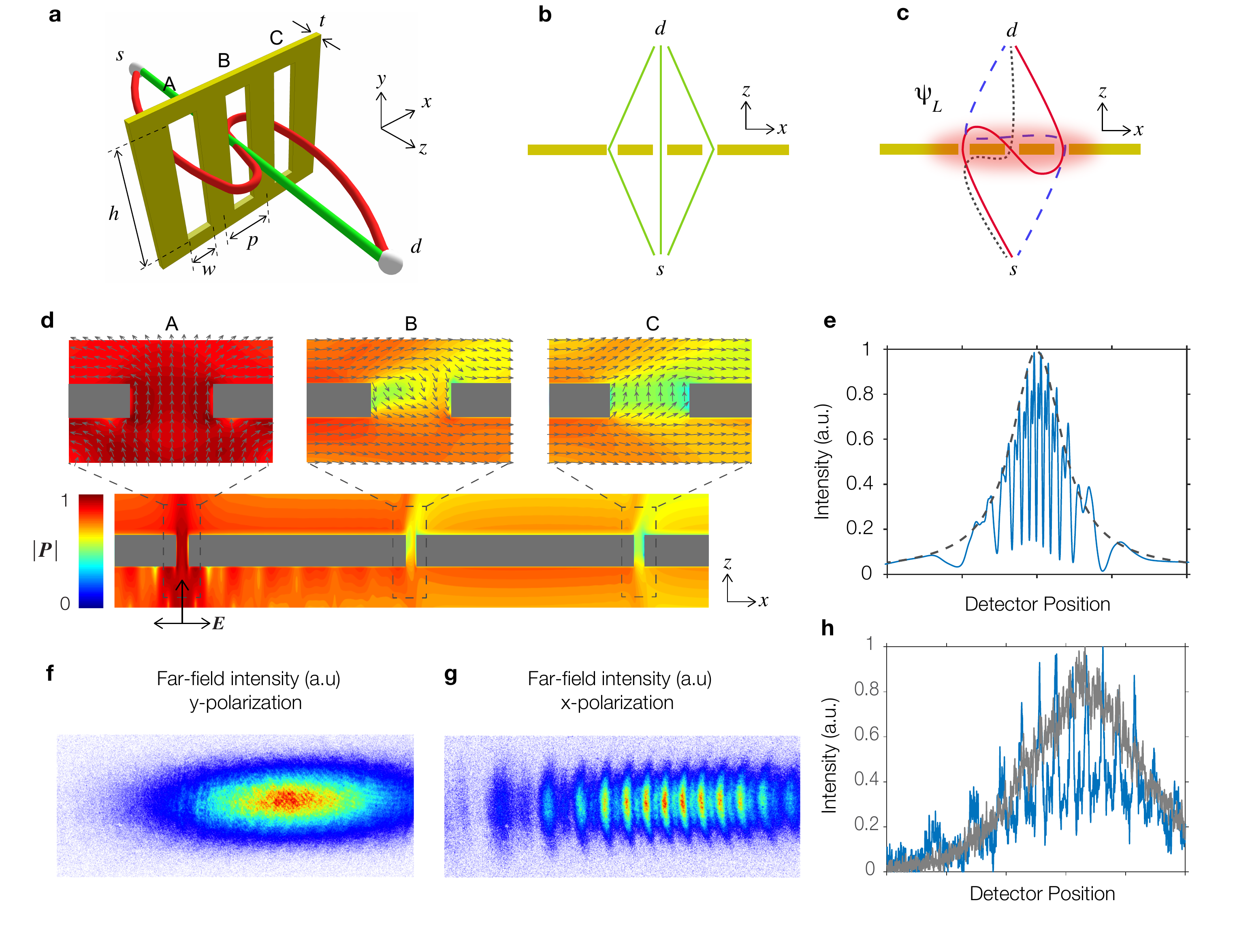}
\caption{{\bf Trajectories of light in a three-slit interferometer.} {\bf a,} The three-slit structure considered in this study. The red path going from point $s$ to point $d$ illustrates a possible looped trajectory of light. {\bf b,} Direct trajectories of light resulting from considering only the first term in Eq.~\eqref{eqn:Ksum}. The widely used superposition principle, i.e. Eq.~\eqref{eq:born}, accounts only for these direct trajectories. {\bf c,} Examples of exotic looped trajectories arising from the higher order terms in Eq.~\eqref{eqn:Ksum}. The red cloud in the vicinity of the slits depicts the near-field distribution, which increases the probability of photons to follow looped trajectories. {\bf d,} Normalized Poynting vector $\mathbf{P}$ in the vicinity of the three slits obtained through full wave simulations at a wavelength $\lambda=810$~nm, using $w=200$~nm, $p=4.6~\mu$m, $t=110$~nm, and $h=\infty$. The simulations consider a Gaussian beam excitation polarized along $x$, and focused onto slit A. The Poynting vector clearly exhibits a looped trajectory such as the solid path in {\bf c}. {\bf e,} Far-field interference patterns calculated under $x$-polarized (solid) and $y$-polarized (dashed) optical excitation. Interference fringes are formed in the far field only when strong near fields are excited ($x$-polarization), and occur from the interference of light following a direct trajectory and a looped trajectory. {\bf f,} Experimental evidence that shows the far-field pattern for a situation in which only one slit is illuminated with $y$-polarized light from an attenuated laser.  {\bf g,} The presence of exotic looped trajectories leads to an increase in the visibility of the far-field pattern. This effect is observed when $x$-polarized light illuminates one of the slits. {\bf h,} The transverse profile of the patterns shown in {\bf f} and {\bf g}. }
\label{fig:fig1}
\end{figure*}

As a concrete example, we consider the situations depicted in Fig.~\ref{fig:fig1}d and \ref{fig:fig1}e. First, we assume a situation in which the incident optical field is a Gaussian beam polarized along the long axis of the slit ($y$ polarization) and focused to a 400-nm spot size onto the left-most slit. For this polarization, surface plasmons are not appreciably excited and the resulting far-field distribution is the typical envelope, with  no fringes, indicated by the dashed curve in Fig.~\ref{fig:fig1}e. This intensity distribution is described by the quantity $\abs{\psi_{\rm A}}^2$. The presented results were obtained through a full-wave numerical analysis based on the finite-difference-time-domain (FDTD) method, on a structure with dimensions $w=200$~nm, $p=4.6~\mu$m, and $t=110$~nm and at a wavelength $\lambda=810$~nm (see Methods). The height of the slit, $h$, was assumed to be infinite. Interestingly, the situation is very different when the incident optical field is polarized along the $x$ direction. The Poynting vector for this situation is shown in Fig.~\ref{fig:fig1}d. This result shows that the Poynting vector predominantly follows a looped trajectory such as that schematically represented by the solid path in Fig.~\ref{fig:fig1}c. The resulting far-field interference pattern, shown as the solid curve in Fig.~\ref{fig:fig1}e, is an example of the interference between a straight trajectory and a looped trajectory. Thus, it is clear that the naive formulation of the superposition principle does not provide an accurate description for the case where near fields are strongly excited.

\vspace{-0.5 cm}
\subsection*{Experimental Results}
\vspace{-0.3 cm}

\begin{figure*}[t]
    \centering
    \includegraphics[width=1\textwidth]{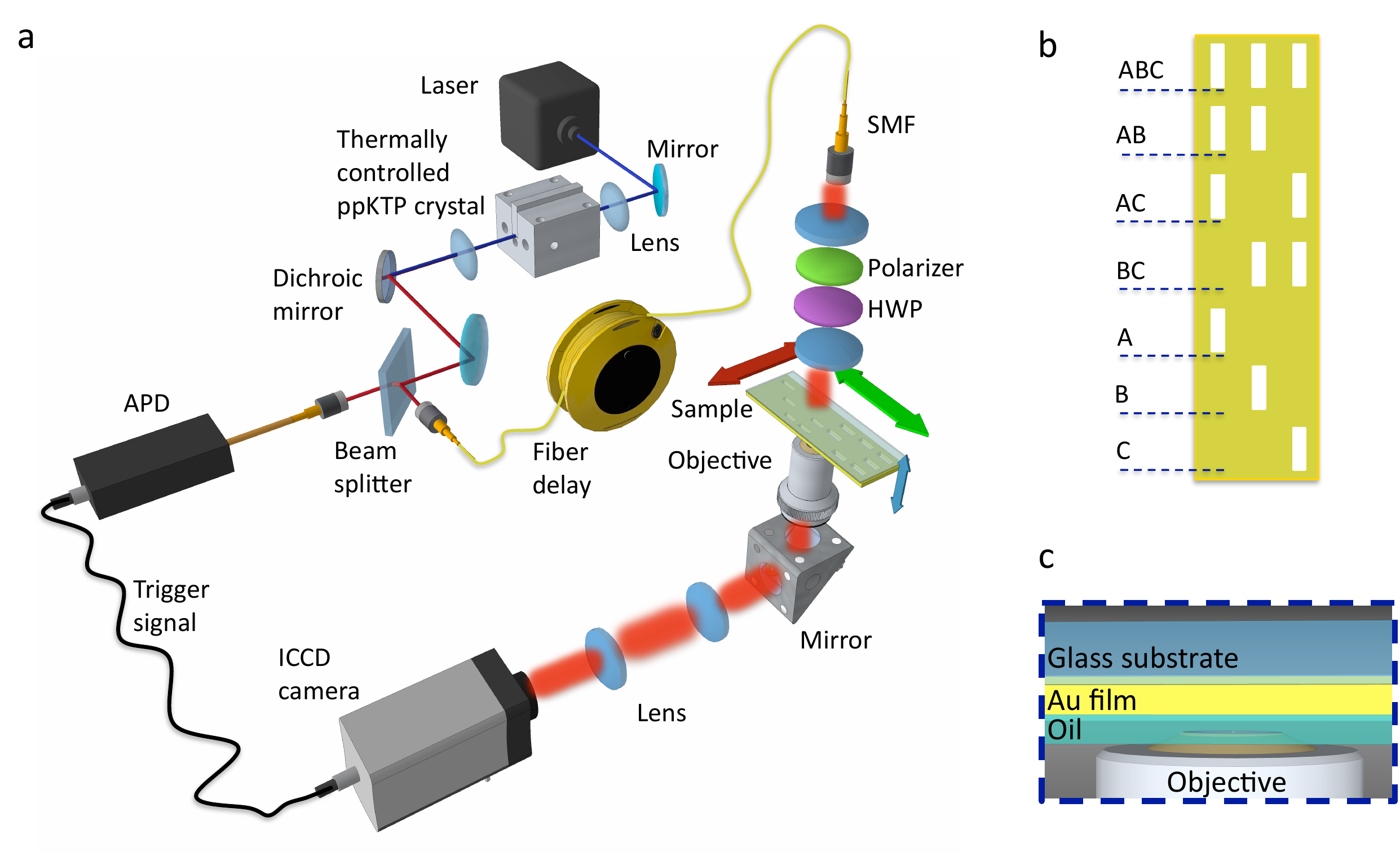} 
    \caption{{\bf Experimental setup utilized to measure exotic trajectories of light}.  {\bf a,} Sketch of the experimental setup used to measure the far-field interference patterns for the various slit configurations.  {\bf b,} The seven different slit arrangements used in our study. This drawing is not to scale; in the actual experiment each slit structure was well separated from its neighbors to avoid undesired cross talk. {\bf c,} Detail of the structure mounted on the setup. The refractive index of the immersion oil matches that of the glass substrate creating a symmetric index environment around the gold film.}
\label{fig:fig2}
\end{figure*}

First, we experimentally verify the role that looped trajectories have in the formation of interference fringes. For this purpose we exclusively illuminate one of the three slits. This experiment is carried out in the setup shown in Fig.~\ref{fig:fig2}a. As shown in Fig.~\ref{fig:fig1}f, no interference fringes are formed when the light illuminating the slit is $y$-polarized. Remarkably, when the illuminating light is polarized along the $x$ direction the visibility of the far-field pattern is dramatically increased, see Fig.~\ref{fig:fig1}g and h. This effect unveils the presence of looped trajectories. In our experiment, the contributions from looped trajectories are quantified through the Sorkin parameter by measuring the terms in Eq.~\eqref{eqn:epsilon}. To this end, we measured the interference patterns resulting from the seven arrangements of slits depicted in Fig.~\ref{fig:fig2}b, thus the illumination field fills each arrangement of slits. In this case, the experiment was carried out using heralded single-photons with wavelength of \SI{810}{nm} produced via degenerate parametric down-conversion (see Methods). The single photons were weakly focused onto the sample, and the transmitted photons were collected and collimated by an infinity-corrected microscope objective (see Fig.~\ref{fig:fig2}c). The resulting interference pattern was magnified using a telescope and recorded using an ICCD camera, which was triggered by the detection event of the heralding photon~\cite{Fickler:SR:2013}. The strength of the near fields in the vicinity of the slits was controlled by either exciting or not exciting surface plasmons on the structure through proper polarization selection of the incident photons.

\begin{figure*}[t]
    \centering
    \includegraphics[width=1\textwidth]{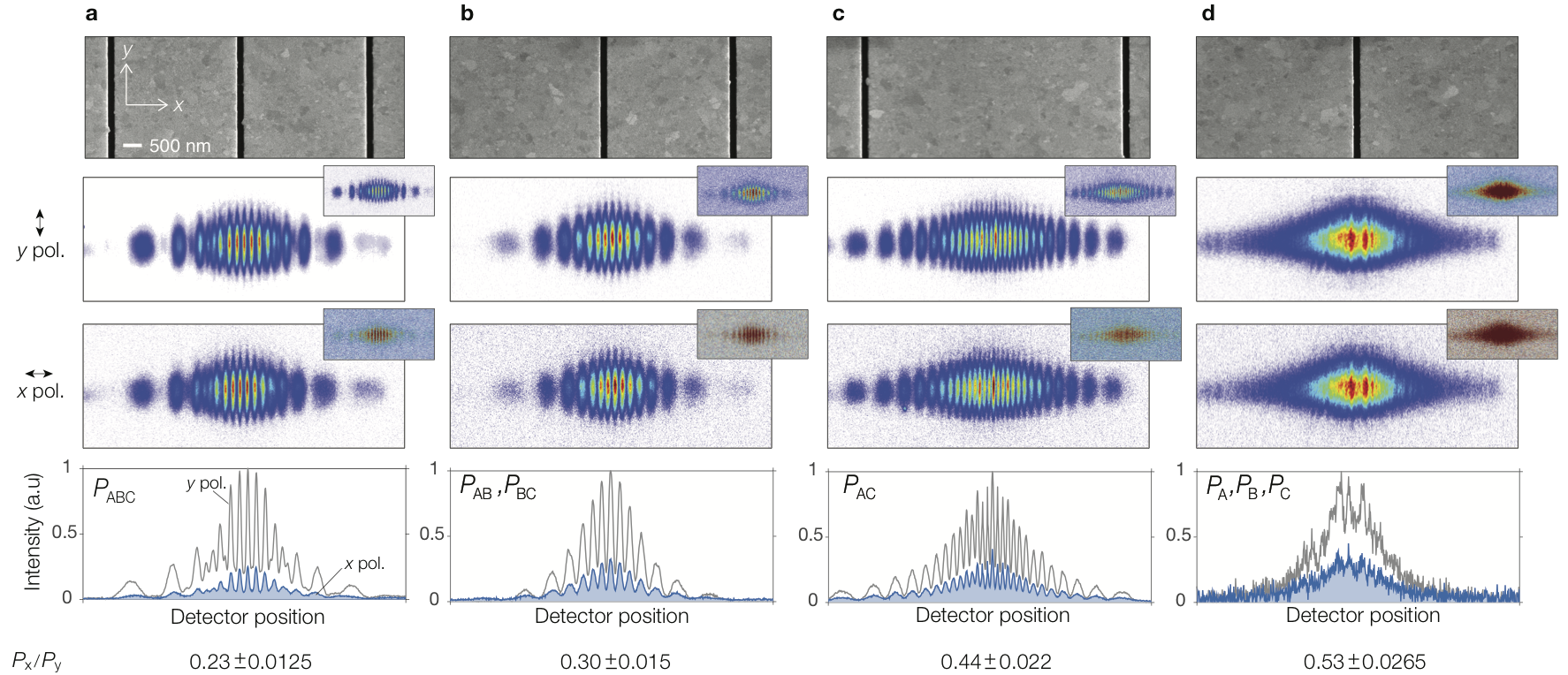}
    \caption{{\bf Experimental results.} {\bf a--d,} Measured interference patterns corresponding to the various probability terms in Eq.~\eqref{eqn:epsilon} (indicated as a label within each panel of the bottom). In this case the illumination field fills each arrangement of slits. The first row shows scanning electron microscope images of the slits used for the measurements. The second and third panels show, respectively, the background-subtracted interference patterns formed when 60 frames, such as those in the insets are added, for the situations in which the probabilities of looped trajectories are negligible (using $y$-polarized illumination), and when such probabilities are increased due to the enhancement of near fields (using $x$-polarized illumination).  Each of the frames shown in the insets was taken with an ICCD camera using heralded single-photons as a source. The bottom show the intensity dependence of the interference pattern measured along a horizontal line on the second and third panels.  The ratio of the average probabilities obtained using $x$-polarized illumination to those obtained using $y$-polarized illumination, $P_x/P_y$, is shown at the bottom. All the measurements are conducted at a wavelength $\lambda=810$~nm, and using structures with dimensions $w=200$~nm, $h=100$~$\mu$m and $p=4.6$ $\mu$m}
\label{fig:fig3}
\end{figure*}

The scanning electron microscope images of the fabricated slits are shown in the first row of Fig.~\ref{fig:fig3}. The dimensions of the slits are the same as those used for the simulation in Fig.~\ref{fig:fig1}, with $h=100~\mu$m being much larger than the beam spot size ($\sim 15~\mu$m). The interference patterns obtained when the contribution from near-field effects is negligible ($y$ polarization) are shown in the second row, while those obtained in the presence of a strong near fields in the vicinity of the slits ($x$ polarization) are shown in the third row. These interference patterns are obtained by adding 60 background-subtracted frames, each of which is captured within a coincidence window of 7 nsec over an exposure time of 160 sec (see insets in Fig.~\ref{fig:fig3}). Only the pattern for $P_{\rm AB}$ is shown in Fig.~\ref{fig:fig3} because $P_{\rm AB}$ and $P_{\rm BC}$ produce nearly identical patterns in the far field, a similar situation occurs for $P_{\rm A}$, $P_{\rm B}$ and $P_{\rm C}$. The bottom panels show detail views of the interference patterns measured along an horizontal line.

Note that the intensities of the interference patterns (i.e., the probability amplitudes) for the two polarizations scale differently for each arrangement of slits. This is shown by the ratios of the position-averaged probabilities, $P_x/P_y$, indicated at the bottom of Fig.~\ref{fig:fig3}. The significant changes in the probabilities obtained with $x$-polarized photons ultimately lead to a value of $\epsilon$ that significantly deviate from zero. This interesting effect is produced by constructive and destructive interference among looped trajectories, whose probability has been increased through the enhancement of the near fields~\cite{SuppMat}. 

\begin{figure*}[t]
\centering
 \includegraphics[width=1\textwidth]{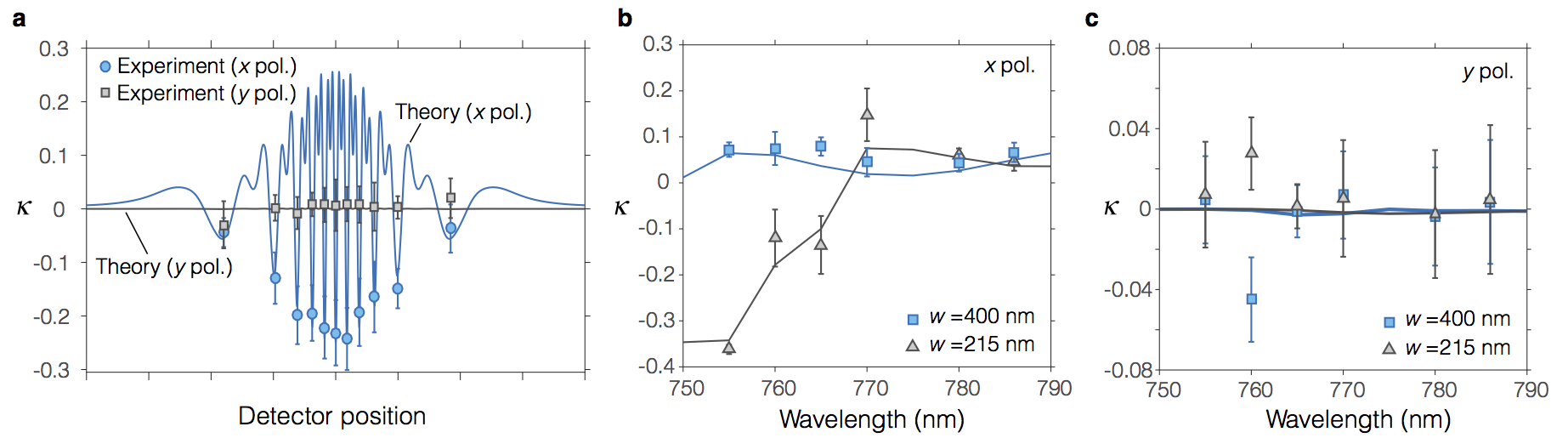}
 \caption{{\bf Quantifying the contribution of looped trajectories through the normalized Sorkin parameter, $\kappa$.} {\bf a} shows numerical and experimental results, for a sample with $w=200$ nm, $h=100$ $\mu$m and $p=4.6$ $\mu$m and an illuminating field consisting of heralded single-photons at a wavelength of $810$ nm. The experimental points are obtained by measuring $\kappa$ at different peaks of the interference patterns shown in Fig. 3. {\bf b} shows theoretical and experimental evidence at the central maximum for different widths and for various wavelengths for an attenuated laser diode, in this case the contributions from looped paths makes the $\kappa$ different from zero. {\bf c} shows a situation in which looped trajectories are not enhanced and consequently $\kappa$ is almost zero. These results confirm that the strengths of looped trajectories can be controlled by engineering the size of the slits and the wavelength of the illuminating field.  These values of  $\kappa$ were measured at the center of the interference pattern. The labels $x$ and $y$ indicate the polarization state of the incident light.}
\label{fig:fig4}
\end{figure*}

We quantify the contribution from the looped trajectories through the normalized Sorkin parameter, defined as $\kappa \equiv \epsilon/I_\text{max}$ with $I_{\rm max}$ being the intensity at the central maximum of the three-slit interference pattern~\cite{Sawant:2014eg}. Both theoretical and experimental values of this parameter are shown in Fig.~\ref{fig:fig4}a. The theoretical values were obtained via FDTD simulations, while the experimental values were calculated from the results in Fig.~\ref{fig:fig3}. Clearly, we observe that when the near fields are {\it not} enhanced, the parameter $\kappa$ is much smaller than the uncertainty associated with our measurements. However, when the near fields are enhanced, $\kappa$ is dramatically increased due to the increased probability for the looped trajectories~\cite{Sawant:2014eg}, enabling the measurement of this parameter despite experimental uncertainties. Taking as a reference the central maximum of the $\kappa$ profile, the experimental results indicate that the contribution of looped trajectories has been increased by almost two orders of magnitude. 

Finally, we show that it is possible to control the probability of photons undergoing looped trajectories by modifying the dimensions of the three slit structure or by changing the wavelength of the optical excitation. Fig.~\ref{fig:fig4}b and \ref{fig:fig4}c show theoretical predictions and experimental data at the central maximum for different slit parameters and wavelengths. These measurements were taken with classical light from a tunable diode laser. Fig.~\ref{fig:fig4}b shows the normalized Sorkin parameter for a situation in which looped trajectories significantly contribute to the formation of interference fringes, whereas Fig.~\ref{fig:fig4}c shows the same parameter for a situation in which near-field effects, and consequently looped trajectories, are negligible. In general, we note that the theoretical and experimental results are in good agreement, with the observed discrepancies being attributed to experimental uncertainties due to imperfections in the fabricated sample and due to the limited dynamic range of the camera.

\vspace{-0.5 cm}
\subsection*{Conclusions}
\vspace{-0.3 cm}

We have demonstrated that exotic looped paths occur as a physical consequence of the near-field component of the wave equation.  As such, it is possible to control the probability of occurrence of such paths by controlling the strength and spatial distribution of the near-fields around the slits. By doing so, we have shown a drastic increase in the probability of photons to follow looped paths, leading to the first experimental observation of such exotic trajectories in the formation of interference fringes. We believe that looped paths can have important implications in the study of decoherence mechanisms in interferometry and to increase the complexity of protocols for quantum random walks, quantum simulators and other algorithms used in quantum computation~\cite{Shadbolt:2014ip, Aspuru:nature:2012}.
 
\subsection*{Methods}
\vspace{-0.2 cm}

\noindent {\bf Sample design:} Full-wave electromagnetic simulations were conducted using a Maxwell's equation solver based on the finite difference time domain method (Lumerical FDTD). The dispersion of the materials composing the structure was taken into account by using their frequency-dependent permittivities. The permittivity of the gold film was obtained from Ref.~\cite{Johnson:PRB:1972}, the permittivity of the glass substrate (BK7) was taken from the manufacturer's specifications, and the permittivity of the index matching fluid (Cargille oil Type B 16484) was obtained by extrapolation from the manufacturer's specification.

\noindent {\bf Sample fabrication:} The glass substrates are standard BK7 cover slips (SCHOTT multipurpose glass D 263$^{\circledR}$ T eco Thin Glass) with a thickness of $\sim 170 \mu$m, polished on both sides to optical quality. The substrate was ultrasonically cleaned for 2 hours in 2\% Hellmanex III alkaline concentrate solution and subsequently rinsed and sonicated in MEK denatured Ethanol and then in demineralized water. The gold films were evaporated directly onto the clean glass substrates with no additional adhesive layer using a Plassys MEB 550S e-beam evaporation system. The growth of the film thickness was monitored in-situ during the evaporation by a water cooled quartz micro-balance. The slit patterns were structured by Ga ion beam milling using a Tescan Lyra 3 GMU SEM/FIB system with a canion FIB column from Orsay Physics. Each slit pattern consisted of $100 \mu$m long slits. While fabricating the different slit sets, proper focusing of the FIB was checked by small test millings and if needed the FIB settings were readjusted accordingly to provide a consistent and reproducible slit quality.

\noindent {\bf Experiment:} We generate single photons by means of heralding a photon by a``partner'' photon detection from a photon pair source. The photon pairs were created in a spontaneous parametric down conversion process using a 2mm-long type-I nonlinear crystal (periodically poled potassium titanyl phosphate (ppKTP)). We pump the crystal with a blue 405nm continuous-wave diode laser ($\sim$200mW), thereby creating degenerate photon pairs at 810nm wavelength. Both photons are passed through a 3nm band-pass filter, coupled into a single-mode fiber and split by a 50/50-fiber beams splitter, which led to a coincidence count rate of approximately 40kHz. The heralding photon is detected with a single-photon avalanche photo diode. Its partner photon is delayed by a 22m long fiber, send through the setup and imaged by an ICCD. The ICCD is operated in the external triggering mode (7ns coincidence gate time), where the heralding detection signal is used as an external trigger, to ensure that only these single photon events are registered~\cite{Fickler:SR:2013}. Note that due to the low coincidence count rate there is only one photon at a time in the experimental setup.
For experiments using a weak laser instead of heralded single photons, the ICCD was operated in the continuous mode, where the intensifier is permanently switched on. 

For the case in which we used single photons, the idler photons are detected by an APD that heralds the detection of signal photons with an ICCD. We used either $y$- or $x$-polarized light which is selected by means of a polarizer and half-wave plate. The beam is weakly focused onto the arrangement of slits that is mounted on a motorized three-axis translation stage that can be displaced in small increments of \SI{60}{nm}.  An infinity-corrected oil-immersion microscope objective (NA=1.4, magnification of 60$\times$, working distance of 100 $\mu$m) was used to collect the light emerging from the slit patterns. The light collected by the objective was then magnified with a telescope and measured by an ICCD camera.

\noindent {\bf Data analysis:} The background subtracted interference patterns were used to determine the magnitudes of $\kappa$ shown in Fig.~\ref{fig:fig4}. In Fig.~\ref{fig:fig4}a, we show the values of $\kappa$, obtained in the single photon regime, for different positions of the detector. The deviation from the theory and the magnitude of the error bars are larger at the edges of the $\kappa$ profile because the signal is low at the edges of the interference patters, which results in a noisier signal. On the other hand, the central maximum of the interference patterns permits a more reliable characterization of $\kappa$. The values of $\kappa$ obtained for classical light as a fuction of the wavelength, shown in Fig.~\ref{fig:fig4}b and ~\ref{fig:fig4}c, were calculated using central maximum of the interference patterns. For these cases, we used only the regions of central fringe having intensities within 70\% of the peak value. The data was then used to obtain the mean value and standard deviations for $\kappa$.





\subsection*{Acknowledgements}
I.D.L., R.W.B. and R.F. gratefully acknowledge the support of the Canada Excellence Research Chairs Program. B. R. was supported by a MITRE Innovation Program Grant. R.F. also acknowledges the support of the Banting postdoctoral fellowship of the Natural Sciences and Engineering Research Council of Canada. U.M. acknowledges the support of DFG under GRK1896. The University of Rochester portion of the work was supported by KBN optics and internal university funds. O.S.M.L and R.W.B. thank P. Milonni, N. Bigelow, N. Vamivakas, Q. Lin and A. Liapis for helpful discussions and comments. We thank A. Gumann and T. Kreller for helping in the preparation of the gold coated glass substrates, and E. Karimi for providing the nonlinear crystal.

\bibliography{Ref}

\end{document}


\title{\textbf{Supplemental Material}\\
    Exotic Looped Trajectories of Photons in Three-Slit Interference}

\author{ 
    Omar S. Maga\~{n}a-Loaiza,$^1$ Israel De Leon,$^{2,3}$  Mohammad Mirhosseini,$^1$ Robert Fickler, $^3$ \\ Akbar Safari,$^3$ Uwe Mick,$^4$  Brian McIntyre,$^1$ Peter Banzer,$^{3,4}$\\ Brandon Rodenburg,$^{5}$ Gerd Leuchs,$^{3,4}$and Robert~W.~Boyd,$^{1,3}$\\
\\ \small{\it $^1$~The Institute of Optics, University of Rochester, Rochester, New York 14627, USA}
\\ \small{\it  $^2$~School of Engineering and Sciences, Tecnol\'{o}gico de Monterrey, Monterrey, Nuevo Le\'{o}n, 64849, Mexico}
\\ \small{\it  $^3$~Department of Physics and Max Planck Centre for Extreme and Quantum Photonics, University of Ottawa,}
\\ \small{\it Ottawa, Ontario, K1N 6N5, Canada.}
\\ \small{\it $^4$~Max Planck Institute for the Science of Light, Guenther-Scharowsky-Str. 1, Erlangen, Germany}
\\ \small{\it  $^5$~Quantum Information Science Group, MITRE, 200 Forrestal Rd. Princeton, NJ 08540, USA} }


\maketitle

\section{Model}
As described in the main text, the path integral formulation of the wave equation propagation kernel is given by
\begin{equation}
    K(\mathbf{r_{\rm 1},r_{\rm 2}}) = \int \mathcal D[x(\mathbf s)]\exp\left(ik\int\dif\mathbf s\right),
    \label{eqn:PathIntegral}
\end{equation}
and can be perturbatively expanded as~\cite{S_Sawant:2014eg}
\begin{equation}
    K = K_1 + K_2 + K_3 + \cdots,
    \label{eqn:Ksum}
\end{equation}
where $K_n$ represents the $n$th application of of the Fresnel-Huygens principle. For instance, considering the propagation through the slits from the source located at $\mathbf{r_s}$ to the detector at $\mathbf{r_d}$, the expression for $K_2$ would be given by
\begin{equation}
\begin{gathered}
    K_2(\mathbf{r_s,r_d}) = \int_\Omega\dif\mathbf{r_1} \int_\Omega\dif\mathbf{r_2} K(\mathbf{r_s,r_1}) K(\mathbf{r_1,r_2}) K(\mathbf{r_2,r_d}),
\end{gathered}
\label{eqn:K2int}
\end{equation}
where integration is over the slit areas $\Omega$. Furthermore, assuming that the slits are much smaller than the wavelength, we can rewrite this equation as a discrete sum over paths. In this case, the expression above becomes
\begin{equation}
K_2(\mathbf{r_s,r_d}) 
    \approx \sum_{j,k} K(\mathbf{r_s,r_j}) K(\mathbf{r_j,r_k}) K(\mathbf{r_k,r_D}),
\label{eqn:DiscretePaths}
\end{equation}
which represents propagation from the source $\mathbf{r_s}$, to slit $j$ at $\mathbf{r_j}$, and then from slit $j$ to $k$ at $\mathbf{r_k}$, and then finally propagation to the detector at $\mathbf{r_d}$.

Each term in Eq.~\eqref{eqn:DiscretePaths} represents a simple propagation solution. Assuming that the slits are illuminated with a planewave normally incident onto the surface of the slits, then the propagation from the source to the screen is
\begin{equation}
    K(\mathbf{r_s, r_A}) = K(\mathbf{r_s, r_B}) = K(\mathbf{r_s, r_C}) = \text{constant},
\end{equation}
and the three straight paths in the (paraxial) far field will be given by
\begin{equation}
\begin{split}
    K(\mathbf{r_B,D}) &\propto \sinc\left(\frac{k_xw}{2\pi}\right)\\
    K(\mathbf{r_A,D}) &= K(\mathbf{r_B,D}) e^{i\phi_f}\\
    K(\mathbf{r_C,D}) &= K(\mathbf{r_B,D}) e^{-i\phi_f},
\end{split}
\end{equation}
where $\theta$ is the far field angle in the ($x,z$) plane ($\theta = 0$ corresponding to propagation in the $z$ direction), $w$ is the slit size, $p$ is the spacing between slits, $k_0=2\pi/\lambda$ is the free-space wavenumber, $k_x = k_0\sin(\theta)$ is the transverse wavenumber, and $\phi_f = pk_x$ is the phase difference due to differences in the distance between each slit and the far field.

For the case at hand, the looped paths involve an enhancement of the near field by surface plasmon modes which we solve for via numerical simulations using the FDTD method. For the simple configuration of slits in a metal screen, it has been shown that this is well approximated by the simple input-output relations~\cite{S_Schouten:2005bl}
\begin{subequations}
\begin{gather}
    K(\mathbf{r_A, r_B}) = 
	K(\mathbf{r_B, r_A}) = 
	K(\mathbf{r_B, r_C}) = 
	K(\mathbf{r_C, r_B}) = a e^{i\phi_{\rm SP}}\\
    K(\mathbf{r_A, r_C}) = 
	K(\mathbf{r_C, r_A}) = b e^{i2\phi_{\rm SP}},
\end{gather}
\end{subequations}
where $a$ and $b$ are the relevant field strengths quantifying the coupling of a free-space to surface plasmon back to free-space mode for neighboring slits and paths with an extra slit between, and $\phi_{\rm SP} = k_{\rm SP}p$ is the phase accumulated along the plasmon path between slits.

\begin{figure}[h]
\centering
\includegraphics[width=0.6\textwidth]{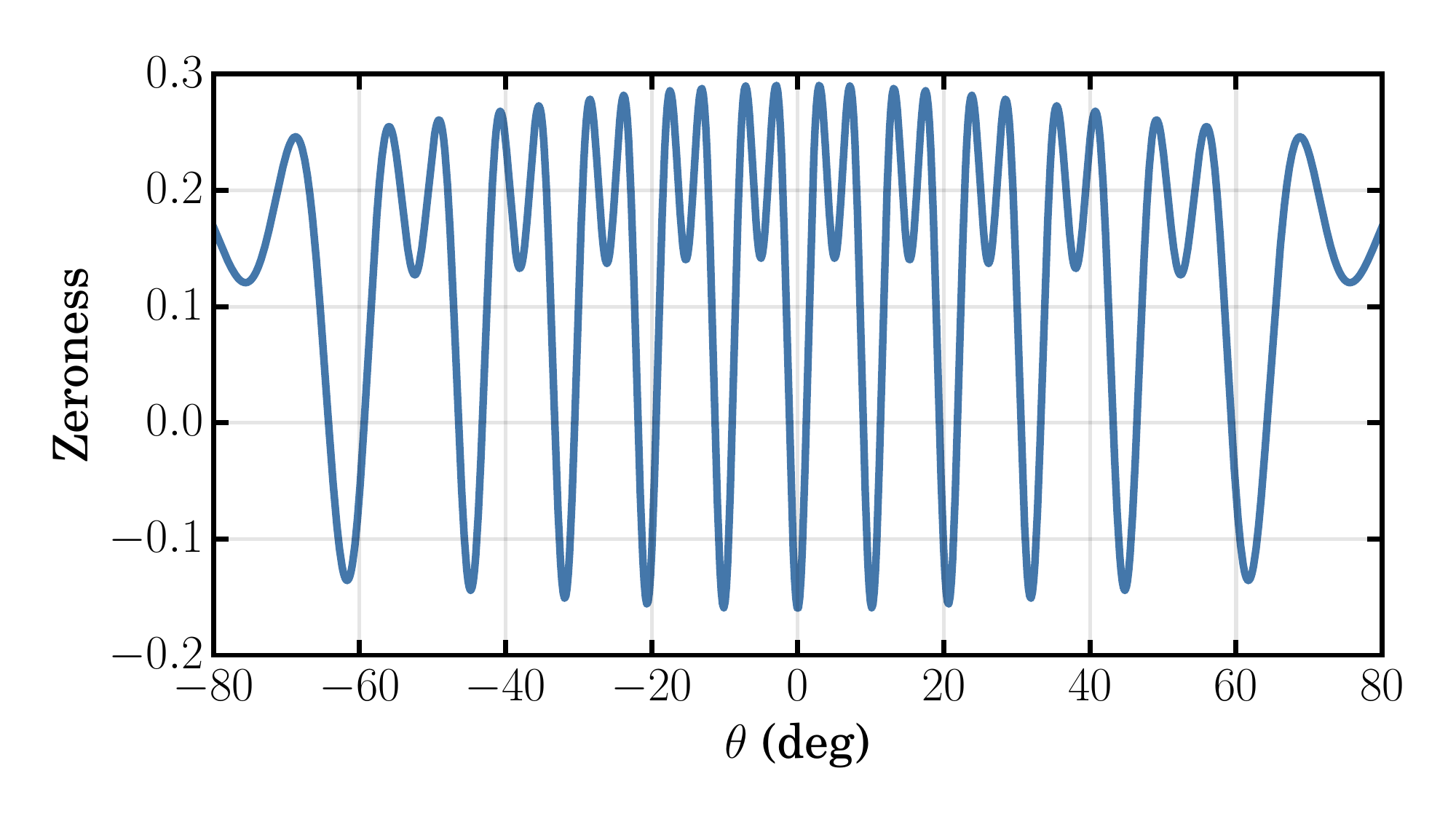}
\caption{Plot of the normalized Sorkin parameter, $\kappa$, assuming experimental parameters similar to those used in the main text.}
\label{fig:3slitplot}
\end{figure}

A plot of the normalized Sorkin parameter, $\kappa$, is shown in Fig.~\ref{fig:3slitplot} assuming the parameters ${p = \SI{4.6}{\micro m}}$, $w = \SI{200}{nm}$, and $\lambda = \SI{810}{nm}$, equivalent to the main paper. This plot was generated using the surface plasmon mode coupling parameters $a = 2b = 0.3$, and $k_{\rm SP} = 1.65k_0$. When this plot is compared with Fig.~4a in the main document, it can be seen that this simple model matches the observed features in both the experiment and full FDTD simulations of the experiment.

\section{Signs of Looped Trajectories}

Since the exotic looped trajectories of light are intimately related to the evanescent component of the fields, the conditions under which the probability of photons following looped trajectories are increased depends on the characteristics of the three slits and the physical processes that take place in their vicinity. The most suitable characteristics for the slits were found by investigating the regimes under which the looped trajectories show a significant role in the formation of interference fringes. We carried out this task by performing a series of FDTD simulations in which different slit sizes, slit separations, and metal thickness were studied.  

\begin{figure}[h]
\centering
\includegraphics[width=1\textwidth]{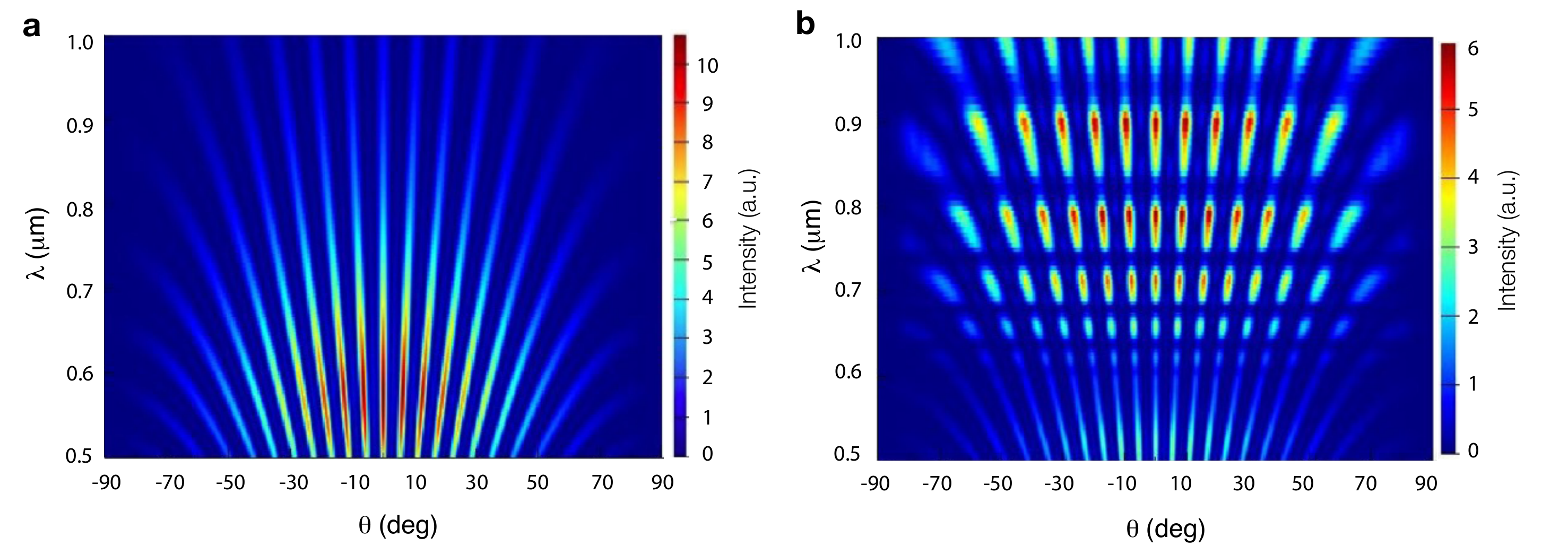}
\caption{The wavelength-dependent intensity distribution of the interference pattern for a three-slit experiment in the absence ({\bf a}) and pretense ({\bf b}) of looped trajectories. Looped trajectories are enhanced by using along the short direction of the slit.}
\label{fig:2slitplot}
\end{figure}

In Fig.~\ref{fig:2slitplot} we show distinct interference structures that unveil the importance of looped trajectories in the formation of interference structures. We plot the intensity as a function of wavelength and detector angles, or positions in the far-field of the three slits. In our calculations, we consider that the slits have a width of $w=150$~nm, the separation between slits is $p=3.6\mu$m, the hight of the slits is infinite, and the three-slit structure is illuminated by a planewave at normal incidence. We show two interference structures produced by the same three-slit structure but under different illumination conditions. When these slits are illuminated with light polarized along the long axis of the slits, surface plasmon modes are not excited in the structure and the interference pattern shown in Fig.~\ref{fig:2slitplot}a is shown in the far field. This pattern is practically identical to that obtained by simply applying the Fourier transform of the three slits. However, this simple experiment shows a striking interference structure when the slits are illuminated with light polarized along the short axis of the slits. In this case, surface plasmon modes are efficiently excited, leading to an increased probability for looped paths, which in turn leads to the significantly different interference pattern shown in Fig.~\ref{fig:2slitplot}b.  In addition, from this pattern it is quite clear how interference effects among looped trajectories lead to a clear wavelength dependence of the experiment. 

The remarkable difference between these two interference structures offered us a clear signature of the presence of looped trajectories. We used this knowledge to decide the characteristics of the slits and the conditions we used when we performed the experiment.


\bibliographystyle{apsrev4-1}
\bibliography{Ref}